\documentclass[aps,prd,onecolumn,groupedaddress,showpacs,nofootinbib,amssymb,notitlepage]{revtex4-1}
\pdfoutput=1
\usepackage[usenames,dvipsnames]{color}
\usepackage[colorlinks=true, linkcolor=BrickRed, citecolor=Blue, urlcolor=Blue, filecolor=Blue]{hyperref}
\usepackage{longtable}
\usepackage{epsfig}
\usepackage{amssymb}
\usepackage{amsmath}
\usepackage{graphicx}
\usepackage{verbatim}
\usepackage{xspace}
\usepackage{hyperref}
\def\aap{A\&A}%
\def\jcap{JCAP}%
\def\mnras{MNRAS }%
\def\prd{Phys.\ Rev.\ D}%
\def\physrep{Phys.\ Rep.}%
\newcommand{\nc}{\newcommand}
\nc{\be}{\begin{equation}}
\nc{\ee}{\end{equation}}
\nc{\ba}{\begin{eqnarray}}
\nc{\ea}{\end{eqnarray}}

\def\bc{\begin{center}}
\def\ec{\end{center}}

\def\to{\rightarrow}

\def\bc{\begin{center}}
\def\ec{\end{center}}

\begin{document}

%------------------------------------------------
\title{Bouncing and emergent cosmologies from ADM RG flows}
%------------------------------------------------

\author{Alfio Bonanno}
\email{alfio.bonanno@oact.inaf.it}
\affiliation{INAF, Osservatorio Astrofisico di Catania, via S. Sofia 78, I-95123 Catania, Italy}
\affiliation{INFN,  Sezione di Catania,  via S. Sofia 64, I-95123, Catania, Italy.}

\author{Gabriele Gionti, S.J.}
\email{ggionti@specola.va}
\affiliation{Specola Vaticana, V-00120 Vatican City, Vatican City State,
\\ Vatican Observatory Research Group, Steward Observatory, The University Of Arizona, 933 North Cherry Avenue, Tucson, Arizona 85721, USA}
\affiliation{INFN, Laboratori Nazionali di Frascati, Via E. Fermi 40, 00044 Frascati, Italy.}

\author{Alessia Platania}
\email{alessia.platania@oact.inaf.it}
\affiliation{INAF, Osservatorio Astrofisico di Catania, via S. Sofia 78, I-95123 Catania, Italy}
\affiliation{INFN,  Sezione di Catania,  via S. Sofia 64, I-95123, Catania, Italy.}
\affiliation{Universit\`a degli Studi di Catania, via S. Sofia 63, I-95123 Catania, Italy}
%\affiliation{Institute for Mathematics, Astrophysics and Particle Physics (IMAPP),Radboud University Nijmegen, Heyendaalseweg 135, 6525 AJ Nijmegen, The Netherlands}

%------------------------------------------------
\begin{abstract}
The Asymptotically Safe Gravity provides a framework for the description of gravity from the trans-Planckian regime to cosmological scales. According to this scenario, the cosmological constant and Newton's coupling are functions of the energy scale whose evolution is dictated by the renormalization group equations. The formulation of the renormalization group equations on foliated spacetimes, based on the Arnowitt-Deser-Misner (ADM) formalism, furnishes a natural way to construct the RG energy scale from the spectrum of the laplacian operator on the spatial slices. Combining this idea with a Renormalization Group improvement procedure, in this work we study quantum gravitational corrections to the Einstein-Hilbert action on Friedmann-Lema\^{i}tre-Robertson-Walker (FLRW) backgrounds.
The resulting quantum-corrected Friedmann equations can give rise to both bouncing cosmologies and emergent universe solutions. Our bouncing models do not require the presence of exotic matter and emergent universe solutions can be constructed for any allowed topology of the spatial slices.
\end{abstract}
\maketitle
%------------------------------------------------

%------------------------------------------------
\section{Introduction}
\label{sect1}
%------------------------------------------------

The standard cosmological model provides a successful and accurate overview of the ``cosmic history'' of our universe \cite{2011BrandenbergerRev}. In this description the existence of an inflationary phase preceding the standard hot Big Bang evolution has become a paradigm. Inflation solves, for instance, the horizon and flatness problems and explains the anisotropies distribution in the Cosmic Microwave Background (CMB) radiation \cite{pla15}. Although the inflationary scenario solves several issues arising in the standard cosmological model, the problem of the Big Bang singularity remains an open question. In fact, assuming the validity of General Relativity, the singularity theorems \cite{1970HP} entail that in a non-closed universe spacetime singularities are inevitable if matter satisfies the standard energy conditions. More specifically, if a non-closed universe undergoes a phase of accelerated expansion, the spacetime must be past geodesically incomplete \cite{2003bgv}.

Alternative scenarios avoiding the initial singularity include bouncing models and the emergent universe scenario. The bouncing universes \cite{1992Mukhanov,1993Brandenberger,2008revnov} are based on the existence of a ``bounce'' at a finite radius separating an initial collapsing phase, in which the universe decreases its spatial volume, and the current expansion phase.
Although bouncing universes avoid the Big Bang singularity, the existence of a nonsingular bounce often entails violations of various energy conditions \cite{cai1,cai2,caipia} or the presence of non-standard matter \cite{linbra,2006bis,2011easson,2011qiu,caiea,2016canfora}.

Another interesting alternative which attempts to replace the standard Big Bang model is the emergent universe scenario \cite{harrison,emergent,ellis03}. According to this model, the universe emerges from an Einstein static state characterized by a non-zero spatial volume with positive curvature. The Einstein static universe is neutrally stable against inhomogeneous linear perturbations \cite{2003barrow}, while it is unstable under homogeneous perturbations \cite{1930eddi}. The latter instability allows the universe to enter the standard inflationary phase. The emergent universe thus avoids the initial Big Bang singularity while preserving the standard energy conditions. On the other hand, it requires a positive-curved spatial geometry. Despite the fact that the recent observational data do not exclude the latter possibility, a spatially flat model is strongly favored \cite{pla15}. 

Modifications of the Einstein theory are expected during the early universe epoch. Although progress in this direction has been hampered by the non-perturbative character of gravity, in recent years the use of Functional Renormalization Group (FRG) techniques has allowed a systematic investigation of gravity under extreme conditions. In fact, a promising approach to construct a consistent and predictive quantum theory for the gravitational interaction is the Asymptotic Safety scenario for Quantum Gravity \cite{Reuter:2012id,Reuter:2012xf,Niedermaier:2006wt,Percacci:2011fr,Litim:2011cp}. As originally proposed by Weinberg \cite{1976W,1979W}, gravity may be a renormalizable quantum theory if the gravitational Renormalization Group (RG) flow runs towards a Non-Gaussian Fixed Point (NGFP) in the ultraviolet (UV) limit. In this case the theory is well defined at all energy scales and the NGFP defines the ultraviolet completion of the gravitational interaction. Using the FRG (non-perturbative) techniques, a UV-attractive NGFP for the gravitational interaction has been found in a large number of truncations schemes \cite{Souma:1999at,martinfrank,Lauscher:2002sq,Codello:2007bd,Machado:2007ea,Falls:2013bv,ast15,Falls:2016msz,Dietz:2012ic,Ohta:2015efa,Ohta:2015fcu,Codello:2006in,Benedetti:2009rx,Groh:2011vn,Gies:2016con,Rechenberger:2012pm,Christiansen:2012rx,Manrique:2011jc,AleFrank1,AleFrank2,2017franktopos}. The key object of the modern FRG \cite{martin,ReuterWetterich,eaa,Benedetti:2010nr} is the so-called effective average action \cite{eaa}, a scale-dependent version of the standard effective action. The evolution of the gravitational action with the RG energy scale is dictated by the functional renormalization group equation (FRGE). The resulting running gravitational couplings can then be used to investigate possible effects of Quantum Gravity in cosmological contexts  \cite{irfp,br02,dou,guberina03,2004reuterw1,2004reuterw2,
guberina05,resa05,boes,br07,weinberg10,cai11,conpe,alfio12,
AlfioAle1,cai13}, see \cite{2017alfr} for a review. For instance, in \cite{Kofinas} it has been shown that the initial singularity may be replaced by a nonsingular bounce. The latter possibility depends on the specific relation between the RG scale and the cosmic time. In fact, the ``cutoff identification'' is not unique and it often depends on the situation at hand. In this regard, an important insight comes from the Arnowitt-Deser-Misner (ADM) \cite{adm,Arnowitt:1962hi} formulation of the FRGE \cite{Manrique:2011jc}. In \cite{Manrique:2011jc} it is argued that, in presence of a foliation structure, the RG scale should be built from the spectrum of the laplacian operator defined on the spatial slices. In the case of a Friedmann-Lema\^{i}tre-Robertson-Walker (FLRW) background this entails a direct connection between the RG scale and the scale factor \cite{Manrique:2011jc}.

Using the scale setting proposed in \cite{Manrique:2011jc}, in this work we study possible cosmological scenarios arising from a quantum-corrected Einstein-Hilbert action evaluated on an FLRW background. The modifications to the Friedmann equations induced by Quantum Gravity allow us to construct both bouncing models and emergent universes. In particular, the resulting models do not require exotic matter fields and are valid for any spatial curvature.

This paper is organized as follows. In Sect. \ref{sect2} we introduce the formalism and derive a quantum-corrected Einstein-Hilbert Lagrangian. Sect. \ref{sect3} provides a detailed constraints analysis of the RG-improved theory obtained in Sect. \ref{sect2}. In particular, we will see how the classical Hamiltonian constraint is modified by Quantum Gravity effects. Sect. \ref{sect4} discusses possible cosmological scenarios arising from our RG-improved model and contains our main results. Finally, a summary of our findings is given in Sect. \ref{sect5}. 

%------------------------------------------------
\section{RG-improved Lagrangian}
\label{sect2}
%------------------------------------------------

Let us consider Einstein-Hilbert action
\be
S =\int d^4 x \sqrt{-g} \, \left\{\frac{R-2\Lambda_k}{16\pi G_k} + \mathcal{L}_m\right\} \label{action}
\ee
where $\mathcal{L}_m$ is the Lagrangian for the matter fields and the subscript $k$ indicates that the gravitational constant and cosmological constant are running quantities whose evolution in the theory space is dictated by the Renormalization Group (RG) equations. Let $\mathcal{L}$ be the total Lagrangian of the system. The specific form of $\mathcal{L}$ can be determined once a specific $(3+1)$ foliation is chosen. In particular, let the spacetime be endowed with a Friedmann-Lema\^{i}tre-Robertson-Walker (FLRW) metric of the type
\be
\label{metric}
ds^2 = -N(t)^2 dt^2 +\frac{a(t)^2}{1-K r^2} dr^2 +a(t)^2 (r^2 d\theta^2 + r^2\sin\theta d\phi^2)\,.
\ee
Here $N(t)$ is the lapse function and $K=-1,0,1$ as usual. In this case the scalar curvature reads 
\be
\label{ricci}
R=\frac{6(-a\dot{a} \dot{N}+ N(a\ddot{a}+\dot{a}^2)+k N^3)}{a^2 N^3}
\ee
where the dot denotes differentiation with respect to the coordinate $t$. 
Let us also assume that the matter field is described by a perfect fluid of energy density $\rho$ and pressure $p$. In this case the relation between $\rho$ and $p$ is parametrized by an equation of state of the type $p=w\rho$, where $w$ is a constant. Therefore the conservation of matter stress-energy tensor ${T^{\mu\nu}}_{;\nu}=0$ with the line element \eqref{metric} fixes the functional form $\rho(a)$ as 
\be \label{rhof}
\rho(a) = m \, a^{-3-3w}
\ee
where $m$ is an arbitrary integration constant. 

In  the construction of the flow equation for the ADM formalism \cite{adm,Arnowitt:1962hi} discussed in \cite{Manrique:2011jc} the infrared cutoff $k$
of the RG transformation is built from the spectrum of the laplacian operator defined on the spatial sections. In particular $k\sim a^{-1}$ in \cite{Manrique:2011jc}. The running Newton's and cosmological constants thus become functions of the scale factor $a(t)$. 

Starting from the action \eqref{action}, a quantum-corrected Lagrangian for a FLRW background can be derived by imposing $k\sim a^{-1}$, using the expression \eqref{ricci} for the Ricci scalar and putting $\mathcal{L}_m=-m N a^{-3w}$ \cite{2014greci}. We thus obtain the following Lagrangian
\begin{equation} \label{lag1}
\mathcal{L}=\, -\frac{3 \, a{\dot a }^{2}}{8\pi N(t)G(a)}+\frac{3 \, a N K}{8\pi G(a)} -\frac{ a^{3}N\Lambda(a)}{8\pi G(a)}-\frac{2 Nm}{a^{3w}} +\frac{3 \, a^{2}{\dot a }^{2}G'(a)}{8\pi N G(a)^{2}}+ \frac{d}{dt} \left(\frac{3 a^2{\dot a }^{2}}{8 \pi NG(a)}\right)
\end{equation}
where $G'(a)$ stands for the derivative of $G$ with respect to $a$ and the total derivative term exactly cancels the York term \cite{1986York} in this case. 

%------------------------------------------------
\section{Constraint analysis}
\label{sect3}
%------------------------------------------------

In analogy to General Relativity, the lapse function $N(t)$ appearing in the Lagrangian~\eqref{lag1} is not a propagating degree of freedom. Therefore, the Hessian determinant associated to the Lagrangian $\mathcal{L}$ vanishes and the dynamics resulting from \eqref{lag1} is degenerate. In classical General Relativity this degeneracy leads to the well-known Hamiltonian constraint \cite{dewitt1967quantum}. 

In the case under consideration the gravitational couplings depend on the degrees of freedom of the system, specifically the scale factor, and the Dirac analysis of constrained systems \cite{dirac1966,anderson1951constraints,bergmann1949non} may give rise to non-standard constraints. In particular, the quantum-corrected Lagrangian \eqref{lag1} acquires new non-classical contributions and  the resulting Hamiltonian constraint could be different from the classical one.
Therefore, before carrying out the analysis of the quantum-corrected cosmological equations, it is of central importance to discuss the constraints arising from the quantum-corrected Lagrangian \eqref{lag1}.

The system under consideration is characterized by one primary constraint. According to the Dirac constraint theory \cite{dirac1966}, the primary constraint $\phi_N (q^i,p_i) \approx 0$ associated to the Lagrangian in eq. \eqref{lag1} is given by
\begin{equation}
p_N=\frac{\partial \mathcal{L}}{\partial {\dot N}}\approx 0\quad\mapsto\quad 
\phi_N(N,a,p_N,p_a)=p_N\approx 0 \;\;.
\label{primary}
\end{equation}
Here, following the Dirac notation, $\approx$ means that the constraint $\phi_N$ is identically zero on the constraint surface $\phi_N=0$ only, namely it holds ``weakly''. The Hamiltonian analysis for degenerate systems requires the canonical Hamiltonian $H_C$ to be defined on the primary constraint surface $M\equiv\{\phi_N=0 \}$ so that
\be
H_C\equiv p_i q^i -\mathcal{L}|_{M}=p_a{\dot a}-\mathcal{L} \;\;.
\ee
Here the momentum $p_a$ associated to the generalized coordinate $a(t)$ is given by
\be
p_a\equiv\frac{\partial \mathcal{L}}{\partial {\dot a}}=-\frac{6 \, a\,{\dot a }}{8\pi N G(a)} \, (\eta(a)+1)
\ee
where $\eta\equiv k\partial_k G_k$, with $k\sim a^{-1}$, defines the ``anomalous dimension'' of Newton's constant as a function of the scale factor
\be \label{anod}
\eta(a)=-\frac{a\,G'(a)}{G(a)} \;\;.
\ee
The canonical Hamiltonian thus reads
\be
H_C=-\frac{2 \pi NG(a)^2p^2_a}{3a(G(a)-aG'(a))}-\frac{3aNK}{8\pi G(a)}+\frac{a^3 \Lambda(a)N}{\pi G(a)}+\frac{2Nm}{a^{3w}}\;.
\label{Ham}
\ee 
As it is well-known, according to the Dirac procedure, one should consider the effective Hamiltonian $\tilde{H}=H_C+\lambda_N\phi_N$, where $\lambda_N$ is a Lagrangian multiplier. Since the constraints must be preserved along the dynamics, we must impose that the dynamical evolution remains on the primary constraint surface
\be
{\dot \phi}_N={\dot p}_N\approx 0\quad\mapsto\quad{\dot p}_N=\big\{p_N,\tilde H \big\}=-\mathcal H\approx 0 \label{impo}
\ee
where $\left \{\,,\right\}$ denotes the Poisson brackets and $\mathcal{H}\equiv N^{-1}{H_C}$ is given by
\be
\mathcal{H}=-\frac{2 \pi G(a)^2p^2_a}{3a(G(a)-aG'(a))}-\frac{3aK}{8\pi G(a)}+\frac{a^3 \Lambda(a)N}{\pi G(a)}+\frac{2m}{a^{3w}} \;\;.
\label{Hamconst}
\ee
The secondary constraint $\mathcal H\approx 0$ thus defines the Hamiltonian constraint associated to the quantum-corrected theory \eqref{lag1}.

Finally, the total Hamiltonian $H_T$ \cite{dirac1966} is obtained from the effective Hamiltonian $\tilde H$ by adding the additional secondary constraints 
\be
H_T=N\mathcal H +\lambda_N\phi_N+\lambda_{\mathcal H} \mathcal H
\label{total}
\ee
$\lambda_{\mathcal{H}}$ being a new Lagrangian multiplier. In analogy to the ADM Hamiltonian analysis of General Relativity \cite{dewitt1967quantum}, it is possible to redefine $H_T$ in the following way
\be
H_T=N\mathcal H +\lambda_N\phi_N\;\;.
\label{ADMHam}
\ee
As it can be easily checked, $\mathcal{H}$ is preserved along the dynamics generated by the total Hamiltonian $H_T$ and therefore no further constraints arise. In particular, the constraints $\phi_N$ and $\mathcal H$ are first class constraints, namely $\{p_N,\mathcal H\}=0$. 

The evolution of the spacetime metric \eqref{metric} is determined by the functional form of the scale factor $a(t)$. The latter can be obtained by solving Friedmann equations once a gauge choice for the lapse function $N(t)$ has been specified. In the context of cosmology the most common choice is $N=1$. This gauge can be formally implemented by introducing the additional constraint $N-1\approx0$. The latter must be preserved along the dynamics generated by the total Hamiltonian $H_T$
\be
\frac{d}{dt}(N-1)=\left\{N-1,H_T\right\}=\lambda_N=0\;\;.
\label{determino}
\ee
This condition fixes the Lagrange multiplier $\lambda_N$ and the total Hamiltonian thus reduces to $H_T=N\,\mathcal{H}$. The constraint $N-1\approx0$ is consistent and hence we fix $N=1$ in the following. 
%Moreover $\{N-1,p_N\}=1$ and $\{N-1,\mathcal{H}\}=0$. Therefore $\mathcal H$ remains a first class constraint, while $p_N$ and $(N-1)$ are of the second class. Therefore we fix N=1 in the following. This remark closes the Dirac constraints analysis.

%------------------------------------------------
\section{Bouncing and emergent cosmologies from Asymptotic Safety}
\label{sect4}
%------------------------------------------------

The Hamiltonian constraint obtained in the previous section provides us with the following RG-improved Friedmann equation
\be
\frac{K}{a^2 H^2}-\frac{8\pi G(a)\, \rho+\Lambda(a)}{3 H^2}+\eta(a)+1=0 \;, \label{1fe}
\ee
where $\eta(a)$ is the quantity introduced in eq.~\eqref{anod} and $\rho(a)$ is given by eq.~\eqref{rhof}. 

Provided $\eta+1\not = 0$, it is useful to put eq. \eqref{1fe} in the following form
\be\label{pot1}
\dot{a}^2=-\tilde{V}_K(a)\equiv-\frac{K+V(a)}{\eta(a)+1}
\ee
where the potential $V(a)$ is given by 
\be \label{pot}
V(a)=\frac{{a}^2}{3}(8\pi G(a)\, \rho+\Lambda(a))
\ee
and the scaling of $G(a)$ and $\Lambda(a)$ is determined by the RG flow. In proximity of the NGFP (early universe) the following approximate solutions of the RG equations can be deduced \cite{br00}
\begin{align}
&G(a)\simeq G_0 \left(1+G_0\, g_\ast^{-1}  a^{-2}\right)^{-1}\\ 
&\Lambda(a)\simeq \Lambda_0 + \lambda_\ast  a^{-2}
\end{align}
where $\lambda_\ast$ and  $g_\ast$ determine the location of the NGFP in the $(\lambda_k,g_k)$ plane, and the constants $\Lambda_0$ and $G_0$  are free parameters which fix the actual RG trajectory emanating from the NGFP. Since $G_0$ and $\Lambda_0$ define the infrared ($k\to0$) limits of the functions $G(a)$ and $\Lambda(a)$, their values should coincide with the observed Newton's constant and cosmological constant, respectively.

\begin{figure}
\begin{center}
\includegraphics[width=0.55\textwidth]{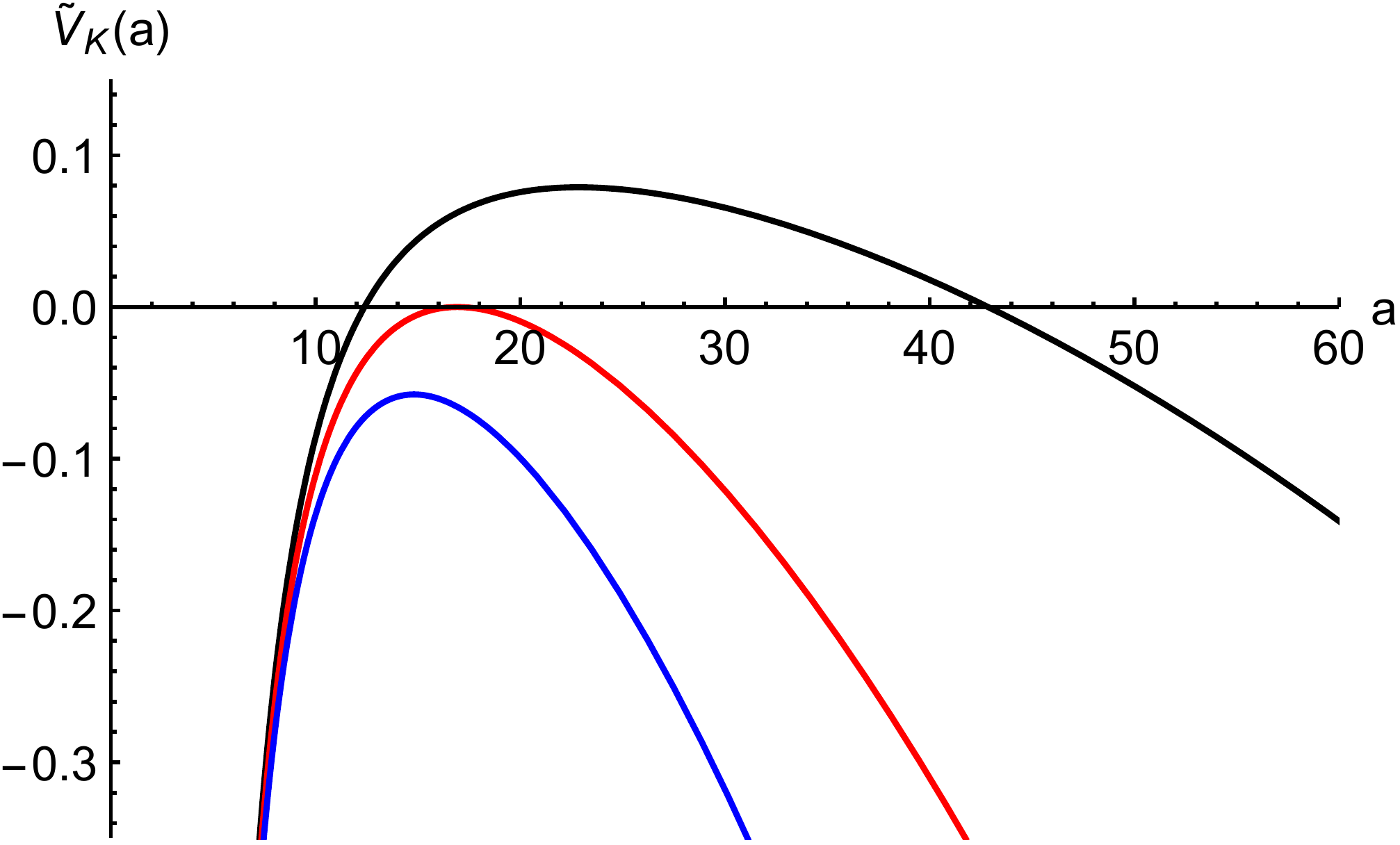}
\caption{The effective potential $\tilde{V}_K(a)$ for a
bouncing universe (black), emergent universe (red), singular universe (blue), for $K=0$, $w=1/3$, $g_\ast=0.1$, $\lambda_\ast=-0.5$ and $m=3$. Black, red and blue correspond to $\Lambda_0=2 \times 10^{-4}$, $\Lambda_0= 8.3 \times 10^{-4}$ and $\Lambda_0=1.5 \times 10^{-3}$ respectively.\label{fig1}}
\end{center}
\end{figure}

We now want to analyze the possible cosmological scenarios arising from the potential \eqref{pot}. As it is apparent from eq.~\eqref{pot1} the allowed regions for the evolution of the scale factor are identified by the condition $\tilde{V}_K(a)\leq0$. In particular, if $\tilde{V}_K(a)=0$ admits real solutions at some $a=a_b>0$, then $\tilde{V}_K(a)$ may give rise to either an emergent universe scenario or a bouncing model. For a general equation of state $p=w\rho$ and spatial curvature $K$, the equation $\tilde{V}_K(a)=0$ implies
\be
%\frac{\Lambda_0}{3 \left(a_0^2-a^2\right)}
\left(a^2+\frac{G_0}{g_\ast}\right) \left(a^2+\frac{\lambda _\ast-3 K}{\Lambda_0}\right)+\frac{8\pi m\,G_0}{\Lambda_0}\,a^{1-3 w}
=0 \;\;. \label{poteqq}
\ee
%and the parameters $l_g$ and $l_k$ are defined as follows
%\be
%l_g=\frac{c\,g_\ast\, \xi ^2\,a_0^2}{\Lambda_0} \qquad\quad 
%l_k=\frac{\lambda _\ast-3 K \xi ^2}{\Lambda_0 \xi ^2}
%\ee
The existence of bouncing solutions depends upon the equation of state and in particular the value of $w$. For sake of simplicity, we restrict ourselves to the case of a radiation-dominated universe, $w=1/3$. For $w=1/3$ eq.~\eqref{poteqq} has (at most) two solutions with non-negative real part. The number of such solutions determines the cosmological scenario arising from $\tilde{V}_K(a)$. For instance, no solutions implies no bounces and the universe has a singularity in the past, at $a=0$. This simple case corresponds to the blue line in Fig.~\ref{fig1}. On the other hand, a bouncing universe is realized when $\tilde{V}_K(a)$ has two different zeros. In particular if $\tilde{V}_K''(a)>0$ the scale factor oscillates between one minimum and one maximum value. On the contrary, if $\tilde{V}_K''(a)<0$, the universe has a bounce at either a minimum \textit{or} a maximum value of the scale factor (black line model in Fig. \ref{fig1}). Only in the former case the initial singularity is avoided.

The most interesting case is the emergent universe scenario.  The key feature of this model lies in a \textit{past-eternal} inflationary phase which naturally follows an initial quasi-static state. The universe thus starts at some minimum scale factor, $a_b>0$. Subsequently it inflates and evolves according to the standard cosmology as predicted by General Relativity. This possibility arises when eq. \eqref{poteqq} has a double zero at some $a_b>0$ such that $\tilde{V}_K''(a_b)<0$, and the special condition $\dot{a}_b=\ddot{a}_b=0$ holds. This case is represented by the red line in Fig. \ref{fig1}.

In the case of an early universe filled with pure radiation, the solutions to eq.~\eqref{poteqq} can be written as follows
\begin{equation}
a_b^2 = -\frac{G_0 \Lambda_0 +g_\ast(\lambda_\ast-3K)}{2 g_\ast \Lambda_0} \pm\sqrt{\left(\frac{G_0 \Lambda_0-g_\ast(\lambda_\ast-3K)}{2 g_\ast \Lambda_0}\right)^2-\frac{8\pi m\,G_0}{\Lambda_0}} \;\;.
\end{equation}
Imposing the uniqueness of this solution basically fixes the integration constant $m$ introduced in eq.~\eqref{rhof}. Moreover, we require $a_b^2$ to be positive, as well as $\tilde{V}''_K(a_b)<0$. The condition to realize a non-trivial ($\tilde{V}_K(a)\leq0$) emergent universe thus reads
\be \label{cond}
\lambda_\ast-3K<-\frac{G_0\Lambda_0}{g_\ast} \;\;.
\ee
In the classical case $\lambda_\ast=0$ and hence, assuming that the infrared values of the cosmological constant $\Lambda_0$ and Newton's coupling $G_0$ are positive, an emergent universe is possible only for positive values of the spatial curvature, $K>0$. On the contrary, the Asymptotic Safety scenario is based on the existence of a NGFP, so that $\lambda_\ast\neq0$. In the latter case the emergent universe can be realized by all allowed topologies of the spatial slices, $K=-1,0,1$, provided that $\lambda_\ast<0$. Remarkably, although $\lambda_\ast>0$ for ``pure gravity'', in the case of gravity-matter systems the value of $\lambda_\ast$ depends on the matter content of the theory. In particular, when gravity is minimally coupled to the matter fields of the Standard Model (or its minor modifications), the gravitational RG flow computed in the ADM-framework has a unique (UV-attractive) NGFP with $\lambda_\ast<0$ \cite{AleFrank2}. {In addition, a cosmological constant which attains negative values in the ultraviolet limit is necessary to ensure the compatibility of Asymptotic Safety with the latest Planck data \cite{AFA}}.

Provided that the condition~\eqref{cond} holds, in the case of an emergent universe eq. \eqref{pot1} reads
\be
\dot{a}^2=-\frac{g_\ast \Lambda_0 \left(a^2-a_b^2\right)^2}{3 \left({G_0}-g_\ast a^2\right)}
\ee
where the minimum scale factor $a_b$ is given by
\be
a_b=\sqrt{-\frac{G_0 \Lambda_0 +g_\ast(\lambda_\ast-3K)}{2 g_\ast \Lambda_0}} \;\;.
\ee
Note that the condition $\tilde{V}_K(a)\leq0$ implies
\be
a^2\geq a_b^2>\frac{G_0}{g_\ast}\;\;.
\ee
Here the quantity $({G_0}/{g_\ast})$ corresponds to the value of $a(t)$ at which the anomalous dimension \eqref{anod} is $\eta=-1$. The value of the minimal length $a_b$ cannot exceed this limit.

In order to understand how the universe evolves in the proximity of $a_b$, i.e. at the very beginning of its evolution, one can linearize the quantum-corrected equation \eqref{pot1} around $a_b$. The resulting approximate equation reads
\be
\dot{a}^2=\frac{4 {g_\ast} a_b^2 \Lambda_0}{3 \left({g_\ast}a_b^2-{G_0} \right)} (a-a_b)^2
\ee
and its general solution is
\be
a(t)=a_b+\epsilon\,\mathrm{exp}\left\{\sqrt{\frac{4 {g_\ast} a_b^2 \Lambda_0}{3 \left({g_\ast} a_b^2-{G_0}\right)}}\; t\right\} \;\;, \label{infla}
\ee
$\epsilon$ being an integration constant. As it is clear from eq. \eqref{infla}, the emergent universe scenario associated with eq. \eqref{poteqq} gives rise to an exponential evolution of the scale factor and no \textit{ad hoc} inflation is needed. In particular, the density parameter can be written as
\be
\Omega-1=\frac{3 \left({g_\ast}a_b^2-{G_0} \right) K}{4 {g_\ast} a_b^4 \Lambda_0}\;e^{-2N_e}
\ee
where the number of e-folds $N_e$ reads
\be
N_e\simeq\mathrm{log}\left(\frac{\epsilon}{a_b}\,\mathrm{exp}\left\{\sqrt{\frac{4 {g_\ast} a_b^2 \Lambda_0}{3 \left({g_\ast} a_b^2-{G_0}\right)}}\; t_e\right\}\right) \;\;,
\ee
$t_e$ being the cosmic time at the inflation exit.

%------------------------------------------------
\section*{Acknowledgement}
%------------------------------------------------

G.G. is grateful to INAF-Catania astrophysical observatory for hospitality.

%------------------------------------------------
\section{Conclusions}
\label{sect5}
%------------------------------------------------

In this work we discussed a class of homogeneous cosmologies consistent with an ADM Renormalization Group evolution.
Our quantum-corrected Friedmann equation provides a new family of bouncing cosmologies which are valid near the basin of attraction of the NGFP and avoid the classical singularity. Emergent universe solutions are also possible depending on the renormalized trajectories around the NGFP. These latter are determined by only two parameters, which are in principle fixed by observations. In particular we showed that our emergent universe models do not depend on the topologies of the spatial sections and do not rely on the presence of exotic matter.

The Dirac analysis shows that the constraint algebra of the quantum-deformed dynamical variables is closed. An interesting question is the generalization of our approach beyond the mini-superspace approximation used in this work. We hope to address this issue in a future work. 

%------------------------------------------------
\bibliography{bbadm}
%------------------------------------------------

\end{document}